\documentclass[twocolumn]{aastex631}
\usepackage{newtxtext,newtxmath}
\usepackage[T1]{fontenc}

\usepackage{graphicx,graphics}	
\usepackage{color,epsfig}
\usepackage{amsmath}	
\usepackage{amssymb}	
\usepackage{psfrag}
\usepackage[T1]{fontenc}
\usepackage{ebgaramond}
\usepackage{newtxmath}

\def\n{\hat{\bm{n}}}
\def\nhat{\hat{\bf n}}

\newcommand{\bfth}{\boldsymbol{\bf{\theta}}}

\newcommand{\kk}{\boldsymbol{k}}
\newcommand{\kp}{\kk_\perp}
\newcommand{\kll}{k_\parallel} 
\newcommand{\xx}{\boldsymbol{x}}
\newcommand{\U}{\boldsymbol{U}}

\newcommand{\gv}{\boldsymbol{g}}

\newcommand{\bfl}{\boldsymbol{\ell}}

\newcommand{\Hi}{H\,\textsc{i}}
\newcommand{\HI}{H\,\textsc{i}~}

\def\tb{{\delta T_{\rm b}}}
\def\ttb{\Delta \tilde{T}_{\rm b}}

\def\a{{\bm{a}}}

\def\v{\bm{v}}

\def\V{\mathcal{V}}
\def\N{\mathcal{N}}
\def\S{{\mathcal S}}
\def\u{{\bf U}} 

\def\pp{\hat{\boldsymbol{\rm p}}}

\defcitealias{Gill_2024_2d3vc}{Paper~I}

\begin{document}

\title{ A Visibility-based 21 cm Bispectrum Estimator for Radio-interferometric Data }

\correspondingauthor{Sukhdeep Singh Gill}
\email{sukhdeepsingh5ab@gmail.com}

\author[0000-0003-1629-3357]{Sukhdeep Singh Gill}
\affiliation{Department of Physics, Indian Institute of Technology Kharagpur, Kharagpur 721 302, India}

\author[0000-0002-2350-3669]{Somnath Bharadwaj}
\affiliation{Department of Physics, Indian Institute of Technology Kharagpur, Kharagpur 721 302, India}

\begin{abstract}

We present a fast and scalable estimator for the binned multi-frequency angular bispectrum (MABS) and the 3D bispectrum (BS) of the redshifted 21 cm signal from radio interferometric observations. The estimator operates on gridded visibilities and leverages the FFT-based acceleration to efficiently compute the MABS and the 3D BS covering all possible triangle configurations. We present the formalism and validate the estimator using simulated visibility data 
for a known input model BS,  considering the Murchison Widefield Array (MWA) observations with a bandwidth of $30.72$ MHz centered at $154.25$ MHz.  We consider two cases, namely, without flagging, and with flagging, which has exactly the same frequency channels flagged as the actual data.  We obtain estimates of the BS for a wide range of triangle shapes covering the scales $0.003 ~\mathrm{Mpc}^{-1}\leq k_1 \leq 1.258  ~\mathrm{Mpc}^{-1}$. The estimated BS shows excellent agreement with analytical predictions based on the input model BS.  We find that the deviations, which are below 20\% even in the presence of flagging, are mostly consistent with the expected statistical fluctuations.   This work paves the way for reliable observational estimates of the 21 cm BS for the epoch of reionization, where the signal is predicted to be highly non-Gaussian.

\end{abstract}



\section{Introduction}


Neutral hydrogen (\Hi\relax), traced by its redshifted 21 cm radiation, has emerged as a promising probe for exploring the large-scale structure and thermal history of the Universe across a broad span of cosmic time \citep{BA5}. Several radio interferometers, such as the Murchison Widefield Array
(MWA)\footnote{\url{https://www.mwatelescope.org/}} \citep{tingay13}, Low Frequency Array (LOFAR)\footnote{\url{https://www.astron.nl/telescopes/lofar/}} \citep{haarlem}, 
and Hydrogen
Epoch of Reionization Array (HERA)\footnote{\url{https://reionization.org/}} \citep{DeBoer_2017} 
 are currently leading the efforts to detect the 21 cm signal from the Epoch of Reionization (EoR). Despite extensive efforts, the auto-correlation detection of the 21 cm signal remains forthcoming, primarily due to contamination from astrophysical foregrounds that are $\sim 4$–$5$ orders of magnitude stronger. Numerous other facilities such as the Canadian Hydrogen Intensity Mapping Experiment (CHIME)\footnote{\url{ https://chime-experiment.ca/en/}} \citep{chime_2022},
MeerKAT\footnote{\url{  https://www.sarao.ac.za/science/meerkat/}} \citep{Cunnington_2023}, and
upgraded Giant Metrewave Radio Telescope 
(uGMRT)\footnote{\url{ https://www.gmrt.ncra.tifr.res.in/}} \citep{uGMRT}  are actively pursuing detection of the post-EoR 21 cm signal, where the foreground contamination is relatively less severe. In a recent work, \cite{Paul_2023} has reported a direct detection of the 21 cm power spectrum using the MeerKAT data at redshifts $z=0.32$ and $0.44$.

The power spectrum (PS) has been the principal statistical tool in observational 21 cm cosmology. This quantifies the variance of brightness temperature fluctuations across different scales. While the PS completely characterizes the Gaussian random fields, it lacks sensitivity to non-Gaussian features. Moreover, it carries no information about phase correlations and is blind to the underlying geometry and topology of the \Hi~ distribution \citep{Bag_2018,Bag_2019}. It is now well established that the redshifted 21 cm signal from both the EoR and the post-EoR is highly non-Gaussian, arising from complex astrophysical and cosmological processes \citep{Bharadwaj2005}. The bispectrum (BS), being the lowest-order statistic sensitive to non-Gaussianity, provides a powerful probe complementary to the PS \citep{Bharadwaj2005}. The BS $B(\kk_1,\kk_2,\kk_3)$ is a function of the closed triangle formed by three Fourier modes $(\kk_1,\kk_2,\kk_3)$, and contains the information on mode coupling. The measurement of the BS can reveal substantial insights that are inaccessible through the PS alone. The utility of the BS has been extensively investigated theoretically and using simulations across various cosmological epochs, including 
the dark ages \citep{pillep2007,cooray2008}, 
cosmic dawn \citep{watk2019, kamran2021}, 
the EoR \citep{shiam2017,suamnm2018,sumanm2020,hutter2020,Saxena_2020,kamran2021, Watkinson_2022,gill_eormulti,raste_2023}, 
and the post-EoR \citep{ali2006,dsarkar2019, durrer2020, cunn2021,Karagiannis_2022,Randrianjanahary_2024}.

The only observational attempt to estimate the 21 cm BS has been made by \citet{trott2019} using the MWA EoR project data, with their analysis restricted to equilateral and isosceles triangles. They demonstrated that thermal noise levels could be reached within 10 hours of observation for large-scale isosceles configurations, and suggested that the BS may be detectable with lesser observational time than the PS.

In recent work \cite{Gill_2024_2d3vc} (hereafter \citetalias{Gill_2024_2d3vc}), we have developed an efficient estimator to compute the angular BS (ABS) $B_A(\ell_1,\ell_2,\ell_3)$ of the sky signal at a single frequency $\nu$ using the radio-interferometric visibility data. In this paper, we extend the framework to multi-frequency observations and present the first visibility-based estimator to compute the multi-frequency angular BS (MABS) $B_A(\ell_1,\ell_2,\ell_3,\nu_1,\nu_2,\nu_3)$. The MABS fully characterizes the 3-point statistics of the $21$ cm signal, assuming the statistical homogeneity and isotropy across the sky. However, it does not require the signal to be ergodic or statistically homogeneous along the frequency axis. Under the additional assumption of ergodicity along frequency, the MABS can be expressed as $B_A(\ell_1,\ell_2,\ell_3,\Delta\nu_1,\Delta\nu_2)$, where $\Delta\nu_1=\nu_1-\nu_3$ and $\Delta\nu_2=\nu_2-\nu_3$. The cosmological 21 cm BS $B(k_{1\perp},k_{2\perp},k_{3\perp},k_{1\parallel},k_{2\parallel})$ is then obtained through the 2D Fourier transform of the MABS $B_A(\ell_1,\ell_2,\ell_3,\Delta\nu_1,\Delta\nu_2)$ \citep{Bharadwaj2005}. In our approach, we first compute the MABS and use it to estimate the 3D BS.   The estimator operates on gridded visibilities and leverages the Fast Fourier Transform (FFT) technique \citep{Sefusatti_thesis,Jeong_thesis,sco2015}, enabling efficient computation of the MABS and the cosmological 21 cm BS across all possible triangle configurations. 

The line of sight (LoS) effects in the observations, such as redshift space distortions (RSD), introduce anisotropy in the BS, making it dependent on the orientation of the $\kk$ vectors relative to the LoS direction. This anisotropy of the BS can be quantified through its multipole moments $B_l^m$, which are the expansion coefficients in the spherical harmonic basis $Y_l^m$ \citep{bharad2020,Mazumdar_2020,gill_2024}. Here, we have presented the formalism to compute all non-zero multipole moments of the BS. However, the validation of the estimator is demonstrated by focusing on the spherically averaged monopole BS.

A brief outline of the paper is as follows. Section \ref{sec:the_ov} presents the theoretical background of the statistics and its parameterization. Section \ref{sec:bsest} introduces the formalism and implementation of the estimator. We discuss our method for validating the estimator in section \ref{sec:validation} and present the results in section \ref{sec:results}. Section \ref{sec:sum} presents the summary and conclusion.

\section{Theoretical Overview}
\label{sec:the_ov}
We consider $\tb(\xx)$ the redshifted 21 cm brightness temperature fluctuations distributed 
across a distant comoving volume spanned by $\xx$. Radio interferometers observe the same signal as $\tb(\nhat,\nu)$ where the unit vector $\nhat$ refers to different directions on the sky, and $\nu$  refers to different observing frequencies.  Note that each frequency corresponds to a different comoving distance that can be calculated from the redshift $z=\left(\frac{1420~\rm{MHz}}{\nu}\right)-1$ of the \HI hyperfine line.  
Here we assume that the angular extent of the observed sky region is sufficiently small, allowing it to be approximated as a two-dimensional (2D) plane. Under this approximation, the unit vector $\nhat$ can be decomposed as  $\nhat = \hat{\mathbf{m}} + \bfth$,  where $\hat{\mathbf{m}}$ represents the unit vector pointing toward the center of the observed field, and $\bfth$ is the 2D angular position vector on the plane of the sky. Consequently, the spatial position $\xx$ can be equivalently described in terms of the coordinates $(\bfth, \nu)$, such that  
\begin{equation}
\xx \equiv (\bfth, \nu) \quad \text{and} \quad \tb(\xx) \equiv \delta T_b(\bfth, \nu).
\label{eq:trans}
\end{equation}

Quantitatively, the perpendicular and parallel components of $\xx$ with respect to $\hat{\mathbf{m}}$ are given by  $\xx_\perp = r \bfth, \quad x_\parallel = r' (\nu - \nu_c),$ where $r$ is the comoving distance corresponding to $\nu_c$, the central frequency of the observation, and $r' = \frac{dr}{d\nu}$ evaluated at  $\nu_c$. 

The field $\tb(\xx)$ can be expressed in terms of its 3D Fourier components $\Delta T_b(\kk)$ as,
\begin{equation}
\Delta T_b(\kk) = \int d^3x ~\exp[ i \kk \cdot \, \xx]\,~\tb(\xx)~ \, ,
\label{eq:3DFT}
\end{equation}
 and also in terms of 2D Fourier components $\Delta \Tilde{T}_b(\bfl,\nu)$,
\begin{equation}
\Delta \Tilde{T}_b(\bfl,\nu) = \int d^2\theta~ \exp[ i \bfl \cdot \, \bfth]\,\delta T_b(\bfth,\nu)  ~ \, , 
\label{eq:TbFT}
\end{equation}  
where $\kk$ and $\bfl$ are the Fourier conjugates of $\xx$ and $\bfth$, respectively. The $\bfl$ may also be interpreted in terms of the angular multipole $\ell=\mid \bfl \mid$. Throughout this work, unless stated otherwise, all integrals extend over the entire range from ${-\infty}$ to $+\infty$.
 
Next, we present the mathematical framework necessary for developing the visibility-based BS estimator. We begin with briefly reviewing the PS before extending the discussion to the BS.

\subsection{MAPS and 3D PS}
 The 3D PS $P(\kk)$ of the brightness temperature fluctuations is defined as,
\begin{equation}
 \langle \Delta T_b(\kk_1) \Delta T_b(\kk_2)  \rangle = {(2\pi)^{3}}\delta_D^3(\kk_1+\kk_2) P(\kk_1),
\label{eq:3dbs}
\end{equation}
where the 3D Dirac delta function $\delta_D^3$ imposes that $\kk_1+\kk_2=0$, and the brackets $\langle \cdots \rangle$ denote an ensemble average over independent realizations of the field. We decompose the vector $\kk \equiv  (\kp, \kll)$  into components $\kp$ and $\kll$ that respectively perpendicular and parallel to the LoS direction $\hat{\mathbf{m}}$, and use 
 $P(\kk_1)\equiv P(\kk_{ 1\perp},k_{1\parallel})$. This decomposition is motivated by two main reasons. First, the observed signal is anisotropic along the LoS direction due to the RSD (\citealt{Bharadwaj_2004}). The decomposition $\kk \equiv  (\kp, \kll)$ is a natural choice to quantify the resulting anisotropy in redshift space.  Although the PS $P(\kk_{ 1\perp},k_{1\parallel})$ exhibits anisotropy along the LoS, it remains isotropic in the plane of the sky ($\kp$), which implies that the PS is independent of the orientation of $\kk_\perp$ and depends only on its magnitude $k_\perp$. Consequently, the PS is a function of two independent parameters, i.e., $P(\kk_1)\equiv P(k_{ 1\perp},k_{1\parallel})$. Secondly, we note that the entire work presented here is focused on the analysis of radio-interferometric observations. Here, the visibilities $\V(\u, \nu)$ are the primary measured quantities, which are recorded as a function of the baseline $\u$ and frequency $\nu$. The baselines $\u$ probe the angular Fourier modes on the sky, and we directly have $\kp = \frac{2\pi \u}{r}$ whereas $k_{\parallel}$ is inferred from $\nu$, the frequency axis. This enable us to reconstruct the cylindrical PS $P(k_{1\perp}, k_{1\parallel})$ from the observed visibilities \citep{Bharadwaj_2001,BA5}.   Importantly, foregrounds, which have a spectrally smooth behavior, are predicted to be localized within the ``foreground wedge'' in the $(k_\perp, k_\parallel)$ plane \citep{adatta10,parsons12,Morales_2012}. The $(k_\perp, k_\parallel)$ modes outside the foreground wedge are expected to be relatively free of foreground contamination, providing the ``21 cm window'' that can be used to estimate the redshifted 21 cm signal. This provides further, very compelling motivation to maintain the distinction between $k_{\perp}$ and $k_{\parallel}$ and consider $P(k_{1\perp}, k_{1\parallel})$ the cylindrical PS. Note that we maintain this distinction for the BS also.

 We now consider the multi-frequency angular PS (MAPS) $C_{\bfl_1} (\nu_1, \nu_2)$ \citep{Datta2007}, defined through 
 \begin{equation}
\begin{aligned}
\langle \ttb(\bfl_1, \nu_1) \ttb(\bfl_2, \nu_2)\rangle=(2\pi)^{2}\delta_D^2(\bfl_1+\bfl_2)C_{\bfl_1} (\nu_1, \nu_2)  \,,
 \label{eq:mabs1}
 \end{aligned}
\end{equation}
which  is a function of the angular multipoles ($\bfl=2 \pi \u =r\kp$) and two frequencies 
$(\nu_1, \nu_2)$.  This assumes the sky signal to be statistically homogeneous and isotropic on the sky, for which it completely quantifies the two-point statistics. Imposing the further assumption that the signal is ergodic along $\nu$, we have 
 $C_{\bfl_1} (\nu_1, \nu_2)\equiv C_{\ell_1} (\Delta\nu)$, where $\Delta\nu=|\nu_2-\nu_1|$. The MAPS is then related to the 3D PS as  \citep{Datta2007},
\begin{equation}
    \begin{aligned}
 &P(k_{ 1\perp},k_{1\parallel}) = 
 r^2 r' \int d(\Delta \nu)   
\exp{[-ik_{ 1\parallel}r'\Delta \nu]} \, C_{\ell_1} (\Delta \nu)~.
\end{aligned}
\label{eq:3dps_maps}
\end{equation}
This relation enables us to compute the 3D PS from the MAPS.

It is possible to directly estimate $C_{\ell_1} (\Delta \nu)$ the MAPS from $\V(\u, \nu)$  the measured visibilities \citep{Ali2008, ghosh1,Choudhuri2014}, and then use Eq.~(\ref{eq:3dps_maps}) to estimate  $P(k_{ 1\perp},k_{1\parallel})$ the 3D PS \citep{Bharadwaj_2019,Pal20}. This approach has been used extensively to analyze data from the uGMRT \citep{elahi_2023_remove,Elahi_2024} and the MWA \citep{chatterjee_2024,Elahi_missing}. 

It is often convenient to express the cylindrical PS as  $P(k_{ 1\perp},k_{1\parallel}) \equiv P(k_1,\mu_1) $ where  $\mu_1=k_{1\parallel}/k_1$, and 
quantify the LoS anisotropy by decomposing $P(k_1,\mu_1) $  into multipoles  using  Legendre polynomials $\mathcal{L}_l(\mu_1)$  \citep{Hamilton_1998}, 
 \begin{equation}
    P_l(k_1)={\int_{-1}^{1}\mathcal{L}_l(\mu_1)~P(k_1,\mu_1)~d\mu_1} \,.
\label{eq:ps_l}
\end{equation}
In the absence of the RSD, we have only the monopole $P_0(k_1)$. The higher-order moments quantify the LoS anisotropy that arises due to the RSD.

\subsection{MABS and 3D BS}
\label{sub:m3d}

The 3D BS $B(\kk_1,\kk_2,\kk_3)$ is defined as,

\begin{equation}
 \begin{aligned}
     \langle& \Delta T_b(\kk_1) \Delta T_b(\kk_2) \Delta T_b(\kk_3) \rangle\\&={(2\pi)^{3}}\delta_D^3(\kk_1+\kk_2+\kk_3) B(\kk_1,\kk_2,\kk_3) \,.
 \end{aligned}
\label{eq:3dbs}
\end{equation}
The 3D Dirac delta function $\delta_D^3(\kk_1+\kk_2+\kk_3)$ imposes the constraint that the three $\kk$ modes form a closed triangle. In the $\kk_\perp$ and $k_\parallel$ coordinates, this constraint yields two conditions $\kk_{1\perp}+\kk_{2\perp}+\kk_{3\perp}=0$ and $k_{1\parallel}+k_{2\parallel}+k_{3\parallel}=0$ \citep{gill_2024}. The first condition implies that the $\kp$ components of three modes form a closed triangle. The BS is invariant under the rotation of $\kp$ around the LoS direction due to isotropy in the plane of the sky, and it depends only on their magnitudes, namely $k_{1 \perp},k_{2 \perp}$ and $k_{3 \perp}$.  The second condition implies that only two $\kll$ components, namely $k_{1\parallel}$ and $k_{2\parallel}$, are independently specified, whereas the third one is determined using $k_{3\parallel}=-k_{1\parallel}-k_{2\parallel}$. Consequently, the BS is a function of only five independent parameters, i.e., $B(\kk_1,\kk_2,\kk_3)\equiv B(k_{1 \perp},k_{2 \perp},k_{3 \perp},k_{1\parallel},k_{2\parallel})$.

  We define the MABS $B_A (\bfl_1, \bfl_2, \bfl_3, \nu_1, \nu_2, \nu_3)$ as,
  \begin{equation}
\begin{aligned}
\langle& \ttb(\bfl_1, \nu_1) \ttb(\bfl_2, \nu_2) \ttb(\bfl_3, \nu_3)\rangle \\&=(2\pi)^{2}\delta_D^2(\bfl_1+\bfl_2+\bfl_3)~B_A (\bfl_1, \bfl_2, \bfl_3, \nu_1, \nu_2, \nu_3) .
 \label{eq:mabs1}
 \end{aligned}
\end{equation}

Assuming ergodicity along the LoS $(\nu)$, and homogeneity and isotropy in the plane of the sky, the MABS is fully characterized by five independent parameters, i.e., $B_A (\bfl_1, \bfl_2, \bfl_3, \nu_1, \nu_2, \nu_3)\equiv B_A (\ell_1, \ell_2, \ell_3, \Delta\nu_1, \Delta\nu_2)$, where $\Delta\nu_1=\nu_1-\nu_3$, $\Delta\nu_2=\nu_2-\nu_3$. The MABS is related to the 3D BS through a 2D Fourier transform \citep{Bharadwaj2005},
\begin{equation}
    \begin{aligned}
 &B(k_{1 \perp},k_{2 \perp},k_{3 \perp},k_{1\parallel},k_{2\parallel}) = 
 r^4 r'^2 \int d(\Delta \nu_1) \int d(\Delta \nu_2)  
\\&\times \exp{[-ir'(k_{1 \parallel}\Delta \nu_1+k_{2 \parallel}\Delta \nu_2)]} \, B_A(\ell_1, \ell_2, \ell_3, \Delta \nu_1, \Delta \nu_2)~.
\end{aligned}
\label{eq:3dbs_mabs}
\end{equation}
Eq.~(\ref{eq:3dbs_mabs}) allows us to compute the 3D BS from the MABS. This is exactly analogous to Eq.~(\ref{eq:3dps_maps}) that allows us to calculate the 3D PS from the MAPS. 
Note that Eq.~(\ref{eq:3dbs_mabs})  is invariant under the transformations $(\Delta \nu_1, \Delta \nu_2)\rightarrow (-\Delta \nu_1, -\Delta \nu_2)$ and $(-\Delta \nu_1, \Delta \nu_2)\rightarrow (\Delta \nu_1, -\Delta \nu_2)$ . Considering the MABS $B_A(\ell_1, \ell_2, \ell_3, \Delta \nu_1, \Delta \nu_2)$ in the $(\Delta \nu_1, \Delta \nu_2)$ plane, with $(\ell_1, \ell_2, \ell_3)$ fixed, we see that it suffices to just consider the upper half-plane i.e. $-\infty \le \Delta \nu_1 \le \infty$ and $0 \le \Delta \nu_2 \le \infty$.  Following \citet{Bharadwaj_2019} that has used the estimated MAPS to determine the 3D PS (Eq.~\ref{eq:3dps_maps}), here we use Eq.~(\ref{eq:3dbs_mabs}) to determine the 3D BS from the estimated MABS. 

Considering the 3D BS $B(\kk_1,\kk_2,\kk_3)$,  in the absence of the RSD, this is independent of how the triangle is oriented i.e.  $B(\kk_1,\kk_2,\kk_3)\equiv B(k_1,k_2,k_3)$, and it 
depends only on the shape and size of the triangle formed by  $(\kk_1,\kk_2,\kk_3)$.  To quantify this dependence, we adopt the parametrization introduced by \cite{bharad2020} that  
uses $k_1$ the largest side ( $k_1\geq k_2\geq k_3$)  to parameterize the size, and two dimensionless parameters  
\begin{equation}
\mu =- \frac{\kk_1 \cdot \kk_2}{k_1 k_2}, \hspace{0.5cm} \, t = \frac{k_2}{k_1} \, ,
\label{eq:shape}
\end{equation}
to parameterize the shape of the triangles. The allowed range of parameter values is restricted to $0.5 \le \mu,t \le 1$ with $ 2 \mu t \ge 1$.

The BS acquires an additional dependence on the triangle’s orientation relative to the LoS direction when the RSD is taken into account. This can be parameterized by a unit vector $\pp$ \citep{bharad2020}, which introduces two additional independent parameters ($p_z,p_x$). The cosines of the angle between the three sides of the triangle and the LoS ($\hat{z}$) are respectively given by 
\begin{eqnarray}
&\mu_{1} & = p_z \,,  \hspace{0.5 cm} \, 
\mu_{2} = -\mu p_z + \sqrt{1-\mu^2}p_x \nonumber \\
&\mu_{3}& = \dfrac{-[(1-t\mu)p_z+t\sqrt{1-\mu^2}p_x]}{\sqrt{1-2t\mu+t^2}}
\label{eq:orient}
\end{eqnarray}


In this parameterization, the BS can be expressed as, $B(\kk_1,\kk_2,\kk_3)\equiv B(k_1,\mu,t,\pp)$, which, as mentioned earlier, requires five independent parameters.  The orientation dependence can be quantified  by decomposing the BS in terms of 
spherical harmonic $Y_l^m(\pp)$, and we have the BS multipole moments \citep{sco1999,bharad2020}, 
\begin{equation}
    \bar{B}_l^m(k_1,\mu,t )=\sqrt{\dfrac{2l+1}{4\pi}}{\int[Y_l^m(\hat{\textbf{p}})]^*~B(k_1,\mu,t,\pp )~d\Omega_{\hat{\textbf{p}}}} \,.
\label{eq:bs_lm}
\end{equation}
We only have the monopole moment $\bar{B}_0^0(k_1,\mu,t)$ in the absence of the RSD. The higher multipole moments \citep{bharad2020,Mazumdar_2020} quantify the effect of the RSD.

\section{The Estimator}
\label{sec:bsest}

\subsection{Formulation}
Visibilities $\V(\u,\nu)$ are the primary quantities measured in radio interferometric observations. Typically, these are measured for a range of baselines $\u$, with components $(\text{u},\text{v})$, and frequencies $\nu$. \citetalias{Gill_2024_2d3vc} considers observations at a single frequency, for which it presents the relation between the ABS $B_A({\ell_1}, {\ell_2}, {\ell_3})$ and the three visibility correlation  $\langle \V(\u_1) \V(\u_2) \V(\u_3)\rangle $. In the present work we generalize this to consider three different frequencies, and present the relation between the MABS and the three visibility correlation 
\begin{equation}
\begin{aligned}
 B_A({\ell_1}, {\ell_2}&, {\ell_3},\nu_1,\nu_2,\nu_3) = \dfrac{3}{\pi\theta_0^2 Q^3} \exp\left [{\pi^2\theta_0^2\Delta U^2/3} \right ] \\& \times\langle \V(\u_1,\nu_1) \V(\u_2,\nu_2) \V(\u_3+\Delta \u,\nu_3)\rangle \,.
\label{eq:S3}
\end{aligned}
\end{equation}
Here $Q=2 k_B/\lambda^2$ is the Rayleigh-Jeans conversion factor   from brightness temperature to specific intensity, and $\theta_0=0.6\theta_F$ (\citealp{Bharadwaj_2001,Choudhuri2014}) where 
 $\theta_F$ is the  FWHM of the primary beam. It is assumed that the frequency range of the observation is sufficiently small that $Q$ and $\theta_{F}$ can be held fixed at the value 
corresponding to $\nu_c$, and we have used $\theta_{F}=24.68\deg$ for the MWA
(\citealt{Line_2018,chatterjee_2023}; \citetalias{Gill_2024_2d3vc}).  
 Here, $\u_1 + \u_2 + \u_3=0$ forms a closed triangle, and $\Delta \u$ is the deviation from a closed triangle configuration. We see that the correlation is the strongest when $\Delta U=0$, and it falls rapidly as $\Delta U$ increases, and there is negligible correlation for $\Delta U \ge (\pi \theta_0)^{-1}$. Note the direct correspondence $\bfl=2 \pi \u$, which has been mentioned earlier. In addition, note that we expect Eq.~(\ref{eq:S3})  to work well at large baselines ($U > 1/\theta_0$). The reader is referred to \citetalias{Gill_2024_2d3vc}  for further details.

\subsection{Implementation}
We have implemented Eq.~(\ref{eq:S3}) to define a visibility-based binned estimator for the MABS. The structure of this section closely follows the approach presented in Section 2 of \citetalias{Gill_2024_2d3vc}. To begin with, we grid the visibility data by introducing a square grid of spacing $\Delta U_g= \sqrt{\ln2}/{\pi\theta_0}$ in the $(\text{u,v})$ plane for each frequency channel, and assign each visibility $\V(\u_i,\nu)$   to the grid point $\u_g$ nearest to $\u_i$ using 
\begin{equation}
\V_{g}(\nu) = \sum^{N_g(\nu)}_{i}\tilde{w}(\u_g-\u_i) \, \V(\u_i,\nu) \, F(\u_i,\nu) ,
\label{eq:vg}
\end{equation}
where $\tilde{w}(\u_g-\u_i)=1$ for the nearest grid point and $=0$ otherwise.  Here 
 the factor  $F(\u_i, \nu)$  accounts for flagging, taking a value of \( 0 \) if the data at baseline $\u_i$  and frequency $\nu$ is flagged, and $1$ otherwise, and $N_g(\nu)=\sum\tilde{w}(\u_g-\u_i)  \, F(\u_i,\nu) $ denotes the number of visibilities that contribute to any grid point $g$ at frequency $\nu$.  


 \begin{figure*}
\centering
\includegraphics[width=1\textwidth]{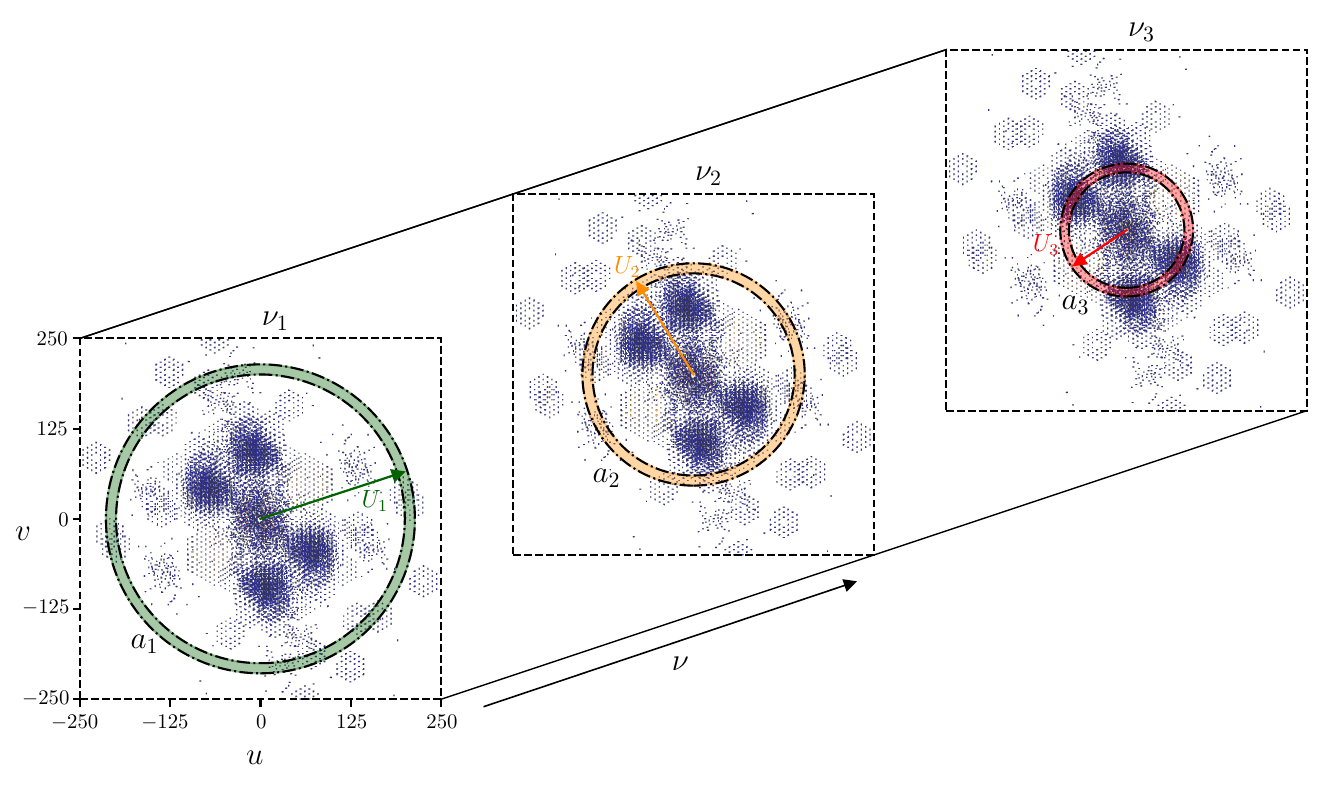}
\caption{The binning scheme used by the MABS estimator. The three $2$D planes correspond to frequency channels  ($\nu_1,\nu_2,\nu_3$).  The scattered dots in each plane show the discrete sampling of (u,v) space (gridded baseline distribution $\u_g$) corresponding to the particular MWA observation considered here.  The $\u_g$ planes are divided into annular rings. Three such rings  (labeled $a_1$, $a_2$ $a_3$)  with average radii $(\ell_1,\ell_2,\ell_3) \equiv 2 \pi \, (U_1,U_2,U_3)$  are shown schematically. 
Note that each ring is on a different plane i.e. a different frequency. The three baselines shown here form a closed triangle $\u_{g1}+\u_{g2}+\u_{g3}=0$. The binned MABS estimator $\hat{B}_A(\ell_1,\ell_2,\ell_3,\nu_1,\nu_2,\nu_3)$ considers the average over all such triangles.  Note that this figure is very similar to Fig.~1 of \cite{gill_2024}, except that the 2D planes there correspond to different values of $k_\parallel$ instead of $\nu$.  }
\label{fig:binV}
\end{figure*}


 We use the gridded visibilities $\V_{g}(\nu)$ to compute the MABS. We have considered the baseline distribution corresponding to a particular pointing of the drift-scan observations presented in \cite{Patwa_2021}. Fig.~\ref{fig:binV} shows the $N_g(\nu)$, the sampling of gridded visibilities for these observations, and also illustrates the binning scheme used in the estimator $-$ a 
 generalization of Fig.~1 of \citetalias{Gill_2024_2d3vc}. Here, each plane corresponds to a different frequency. The computational complexity of directly evaluating the correlations for all possible triplets of grid points that form closed triangles scales as $\mathcal{O}(N_c^3N_t^4)$, where $N_c$ is the total number of frequency channels and $N_t$ is the number of grid points in each dimension.  
 The computational cost can be reduced to $\mathcal{O}(N_c^3N_t^2\log{N_t^2})$ by utilizing the FFT techniques (\citealt{sefu2006, sco2015,shaw21}; \citetalias{Gill_2024_2d3vc}). 
 
 Here, for each frequency channel, we divide the $\u$ planes into $N_r$ concentric annular rings. Three such rings,  labeled  $(a_1,a_2,a_3) $ with  mean radii $(U_1,U_2,U_3)$  respectively,  are illustrated in  Fig. \ref{fig:binV}. Considering any ring $a_m$ at frequency $\nu$, we define,
\begin{equation}
    D(\ell_m,\nu,\bfth)=\sum_{g\in a_m} W_{g}(\nu) \V_g(\nu) \exp(-i  \bfl_g \cdot\bfth)~, 
    \label{eq:iFT}
\end{equation} 
which is the inverse Fourier Transform of $\V_g(\nu)$ restricted to the annular ring $a_m$, with mean radius $\ell_m=2 \pi U_m$. 
We have introduced weights $W_g(\nu)$ for the gridded visibilities, which can be adjusted to optimize the signal-to-noise ratio of the estimated MABS.
Here, we have not included the system noise contribution, and we use $W_g=N_g^{-1}(\nu)$  and $0$ for the filled and empty grids, respectively. 
Similarly,
\begin{equation}
    I(\ell_m,\nu,\bfth)=\sum_{g\in a_m}  W_g(\nu) N_g(\nu) \exp(-i\bfl_g \cdot\bfth)~, 
    \label{eq:iFT_I}
\end{equation} 
 is the inverse Fourier Transform of $ W_g(\nu) N_g(\nu)$ restricted to the annular ring $a_m$.  

We consider the combination of three annular rings  $( a_1, a_2, a_3)$  respectively located in the channels $(\nu_1,\nu_2,\nu_3)$ and define the binned estimator for the MABS (\citetalias{Gill_2024_2d3vc}),
\begin{equation}
\begin{aligned}
    &\Hat{B}_A(\ell_1,\ell_2,\ell_3,\nu_1,\nu_2,\nu_3)
\\&= \dfrac{1}{\mathcal{A}}  \dfrac{\sum\limits_{\bfth} D(\ell_1,\nu_1,\bfth) D(\ell_2,\nu_2,\bfth)  D(\ell_3,\nu_3,\bfth)}{\sum\limits_{\bfth} I(\ell_1,\nu_1,\bfth) I(\ell_2,\nu_2,\bfth)  I(\ell_3,\nu_3,\bfth)} ~.
\end{aligned}
\label{eq:estimator_t}
\end{equation} 
 Here, $\mathcal{A}=\pi\theta_0^2Q^3/3$ is a normalization constant (Eq.~\ref{eq:S3}).

The estimator considers all closed triangles $\u_{g1}+\u_{g2}+\u_{g3}=0$ such that $(\u_{g1},\u_{g2},\u_{g3})$ are within the annular rings $(a_1,a_2,a_3)$ at corresponding frequencies $(\nu_1,\nu_2,\nu_3)$ respectively, for which one triangle is illustrated in Fig. \ref{fig:binV}.  The binned MABS estimator $\hat{B}_A(\ell_1,\ell_2,\ell_3,\nu_1,\nu_2,\nu_3)$ 
(Eq.~\ref{eq:estimator_t}) provides the weighted average over all such closed triangles. 
 Further, we assume the signal to be ergodic along frequency, and we collapse the MABS to express it as a function of frequency separations 
 $\hat{B}_A(\ell_1,\ell_2,\ell_3,\Delta\nu_1,\Delta\nu_2)$. This is related to the 3D BS  through a 2D Fourier transform on the $(\Delta \nu_1,\Delta \nu_2)$ plane (Eq.~\ref{eq:3dbs_mabs}). Here we implement a 2D DFT,  
\begin{equation}
    \begin{aligned}
 &\hat{B}(k_{1 \perp},k_{2 \perp},k_{3 \perp},k_{1\parallel a},k_{2\parallel b}) = 
  (r^2 r'\Delta\nu_c)^2 \sum\limits_{{n,m=-N_c}/{2}}^{{N_c}/{2}} 
\\&\times \exp{[-2\pi i(na+mb)/N_c]} \, \hat{B}_A(\ell_1, \ell_2, \ell_3, n\Delta \nu_c, m\Delta \nu_c)~,
\end{aligned}
\label{eq:dft}
\end{equation}
where $k_{1\parallel a} =  2\pi a/ (r' B_{\rm bw})$, $\kk_{1\perp}={\bfl_1}/{r}$, etc., and $N_c$, $\Delta \nu_c$  and $B_{\rm bw}$ respectively correspond to the total number of frequency channels, channel width and total bandwidth.  

Finally, we evaluate the BS multipoles by implementing a discrete version of Eq.~(\ref{eq:bs_lm}) (refer to Eq. (13) of \cite{gill_2024}). In this study, we have restricted our analysis to the monopole moment only.

The algorithm presented is inherently parallel. The 2D FFT operations in Eqs.~(\ref{eq:iFT}) and (\ref{eq:iFT_I}), corresponding to different rings and frequency channels, can be independently parallelized across multiple threads. Likewise, the estimator in Eq.~(\ref{eq:estimator_t}) can be evaluated in parallel by assigning each combination of three rings and channels to separate threads, allowing efficient large-scale computation.

\section{Validating the estimator}
\label{sec:validation}
We validate the estimator by simulating a non-Gaussian sky signal $\delta T_{\rm b} (\xx)$ for which the analytical expressions for the BS and the MABS are known. We start from a Gaussian random field $\delta T_{\rm G} (\xx)$ generated using an input model PS $P(k)=k^{-2}$.  The non-Gaussian random field $\delta T_{\rm b} (\xx)$ is obtained using a local non-linear transformation,
\begin{equation}
\delta T_{\rm b} (\xx)= \delta T_{\rm G} (\xx) + {f_{\rm NG}}\,(\delta T^2_{G} (\xx)- \sigma_{ T}^2) \,, 
\label{eq:mb1}
\end{equation}
where the dimensionless parameter $f_{\rm NG}$ controls the level of non-Gaussianity, and $\sigma_{ T}$ is the standard deviation of $\delta T_{\rm G} (\xx)$. The analytical expression for the BS of $\delta T_{\rm b} (\xx)$ calculated to the first order in $f_{\rm NG}$ is,
\begin{equation}
\begin{aligned}
    &B_{\rm Ana}(k_1,k_2, k_3) \\&= 2\,f_{\rm NG}\big[{P(k_1)}\,P(k_2)+{P(k_2)}\,{P(k_3)}+ {P(k_3)}\,{P(k_1)}\big] \,,
\end{aligned}
\label{eq:bsana}
\end{equation}
which is valid for $f_{\rm NG}\sigma_{ T}\ll1$. Here we have used $f_{\rm NG}=1$, for which $f_{\rm NG}\sigma_{ T}\approx 0.25$ and the BS estimated from the simulated sky signal was found to be consistent with the predictions of Eq.~(\ref{eq:bsana}). The analytical expression for the MABS can be obtained by performing a 2D inverse Fourier transform of the 3D BS given in Eq.~(\ref{eq:bsana}), effectively inverting Eq.~(\ref{eq:3dbs_mabs}).

We have used the baseline distribution corresponding to a particular pointing of the drift scan observations of the MWA presented in \citet{Patwa_2021} and also analyzed in \cite{chatterjee_2023,chatterjee_2024}; \citetalias{Gill_2024_2d3vc}. The observations consist of $N_c=768$ frequency channels, each of width $\Delta \nu_c=40$ kHz, resulting in a total bandwidth of $B_{\rm bw}=30.72$ MHz. The central frequency of the observation is $\nu_c = 154.2$ MHz. The data exhibits a periodic pattern of missing frequency channels, which we now describe. The entire bandwidth is divided into $24$ sub-bands, each comprising of $32$ frequency channels. In each sub-band, the four edge channels at both ends and one central channel are flagged, resulting in a periodic flagging pattern with a frequency spacing of $1.28$ MHz. In addition to this systematic flagging, data may also be flagged across different baselines due to radio frequency interference (RFI). The baselines in the data are mostly  $(\sim 99 \%)$   within $U =250$ \citep{chatterjee_2023}, which  corresponds to $\ell =1570$ or an angular scale of $0.115^{\circ}$. Here, we have simulated the sky signal on a 3D box with $2048$ grids in each direction. Considering the $\xx_\perp$ planes, they span  $50.96^{\circ}$ that is $\sim 2.12$ times larger than $\theta_{F}$, with a resolution of $0.025^{\circ}$.  We have chosen this large dynamical range of angular scales to avoid abruptly cutting off the simulated signal at either large or small angular scales. The grid spacing in the $x_\parallel$ direction is $0.68$ Mpc, which corresponds to $40$ kHz that matches the MWA channel width. We have used Eq.~(\ref{eq:trans}) to convert the simulated signal from $\tb(\xx)$ to $\tb(\bfth,\nu)$, and we calculate the simulated visibilities $\V(\u,\nu)$  using exactly the same prescription as in  \citetalias{Gill_2024_2d3vc}.  The simulated data incorporates exactly the same flagging as the actual data, and in our analysis, we consider both, with and without flagging. 
Note that our simulations do not include the variation in the primary beam pattern and baseline with frequency. In principle, we should use a spherical sky to perform these simulations (e.g., \citealp{chatterjee_2023}), however, the simulations are significantly faster if we use the flat sky approximation adopted here. Further, the subsequent analysis is restricted to the range $20 \le \ell \le 1570$ where we expect the flat sky approximation to hold \citep{Datta2007}.

We grid the visibilities according to Eq.~(\ref{eq:vg}), with grid spacing of $\Delta U_g= \sqrt{\ln2}/{\pi\theta_0}\approx 1$. Here, instead of correlating the visibilities $\V(\u,\nu)$ at three baselines that form a closed triangle (Eq.~\ref{eq:S3}), we estimate the MABS by correlating  $\V_g(\nu)$ at three grid points that form a closed triangle. The different baselines that contribute to the three grid points generally do not form closed triangles. 
The relatively small grid spacing used here ensures that the factor $e^{-(\pi^2\theta_0^2\Delta U^2/3)}$ that arises in Eq.~(\ref{eq:S3}) 
due to this does not fall much below $1$. We find that this factor has a value of $0.89$ for a typical value of $(\Delta U)^2=(\Delta U_g)^2/2$. To reduce computational load, we collapse four consecutive frequency channels, reducing the volume from $768$ to $192$ channels. This results in a coarser frequency resolution of 160 kHz while preserving the total bandwidth of 30.72 MHz and maintaining the essential spectral structure of the data.

We have divided the $(\text{u},\text{v})$ plane (Fig.~\ref{fig:binV}) 
 into $22$ concentric annular rings of varying width, with a single ring of width 4  spanning radius 1 to 5 (in grid units), 
  nine equally spaced rings between radii $5$ and $50$, five between 50 and 100, five between 100 and 200, and two between 200 and 250. We have used Eq.~(\ref{eq:estimator_t}) to estimate  $\hat{B}_A(\ell_1,\ell_2,\ell_3,\nu_1,\nu_2,\nu_3)$ for every possible combination of three annular rings and three frequencies. The  widths of the annular rings increase progressively with $U$ to account for the fact that the expected signal $B_A(\ell_1,\ell_2,\ell_3,\nu_1,\nu_2,\nu_3)$, and the baseline number density, both decrease with increasing $U$.   Choosing a larger number of finer rings would provide estimates of the MABS at small intervals of $(\ell_1,\ell_2,\ell_3)$, at the cost of increasing the computation time, while choosing a smaller number of coarser rings would have the opposite effect. 
  
  
  We note that the system noise contribution to the different visibilities is uncorrelated, and this does not contribute to the expectation value of the estimator.  We have used $250$ statistically independent realizations of the non-Gaussian random field to obtain reliable estimates of the ensemble averaged MABS  ${B}_A(\ell_1,\ell_2,\ell_3,\nu_1,\nu_2,\nu_3) =\langle \hat{B}_A(\ell_1,\ell_2,\ell_3,\nu_1,\nu_2,\nu_3) \rangle$ and the 3D BS $B(k_{1 \perp},k_{2 \perp},k_{3 \perp},k_{1\parallel},k_{2\parallel})= \langle \hat{B}(k_{1 \perp},k_{2 \perp},k_{3 \perp},k_{1\parallel},k_{2\parallel}) \rangle$. The analysis is carried out separately considering the data with and without the missing (flagged) frequency channels, and the results are discussed in the following section.


\section{Results}
\label{sec:results}

 \begin{figure*}
\centering
\includegraphics[width=1\textwidth]{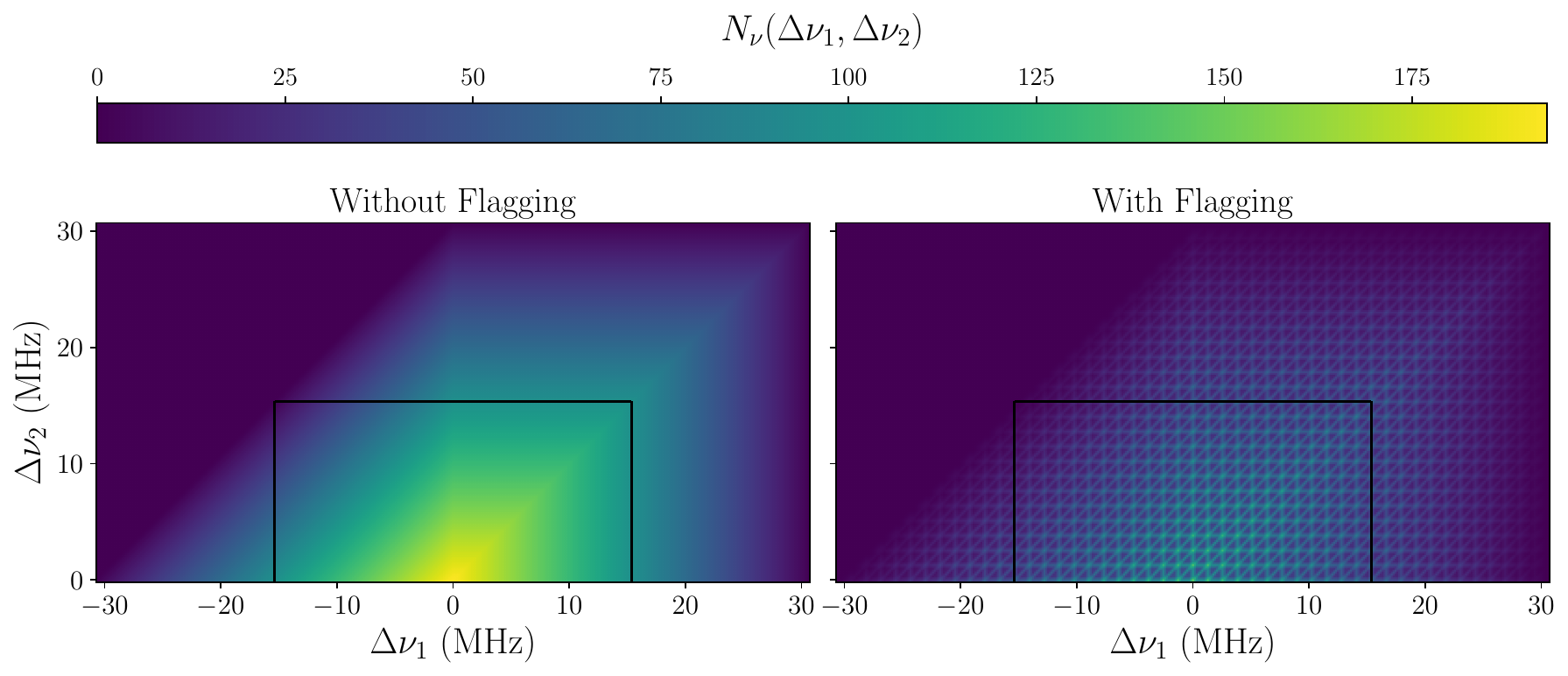}
\caption{ The sampling  $N_\nu(\Delta\nu_1,\Delta\nu_2)$, which counts the number of frequency channel triplets ($\nu_1,\nu_2,\nu_3$) corresponding to each set of frequency separations  $(\Delta\nu_1,\Delta\nu_2)$ . The left panel shows $N_\nu(\Delta\nu_1,\Delta\nu_2)$ without flagging, while the right panel shows the same with flagging i.e. it incorporates the periodic pattern of missing channels present in the MWA data. In both panels, the region where 
$|\Delta \nu_2 - \Delta \nu_1| > B_{\rm bw}$ is not sampled at all.  The black line encloses the region that we have used for the 2D FFT (Eq.~\ref{eq:dft}) to evaluate the 3D BS. This is 
the largest regular region that avoids the frequency separations that are not sampled.  
Note that there are no missing frequency separations $(\Delta\nu_1,\Delta\nu_2)$ introduced due to the flagging.}
\label{fig:Nv}
\end{figure*}

Fig.~\ref{fig:Nv} shows $N_\nu(\Delta\nu_1,\Delta\nu_2)$, which quantifies the sampling of  
$(\Delta\nu_1,\Delta\nu_2)$  as $(\nu_1,\nu_2,\nu_3)$ span all possible combinations. 
As noted in sub-section~\ref{sub:m3d}, it suffices to only show the upper-half plane, and we have added the values from the lower-half plane to the corresponding points on the  
upper-half plane. The sampling  $N_\nu(\Delta\nu_1,\Delta\nu_2)$ corresponding to our simulation 
will be reflected in the estimated values of ${B}_A(\ell_1,\ell_2,\ell_3,\Delta\nu_1,\Delta\nu_2)$, and we expect smaller statistical fluctuations in regions where the sampling is high. We first consider the left panel
which does not incorporate the effect of flagging. We see that  $N_\nu(\Delta\nu_1,\Delta\nu_2)$ peaks at $(\Delta \nu_1,\Delta \nu_2) = (0,0)$  where $N_\nu=N_c$, and  
falls approximately linearly with $|\Delta \nu_1|$ and $|\Delta \nu_2|$ as there exist fewer valid frequency triplets at large separations. The decline is slowest along $\Delta \nu_2=\Delta \nu_1$, and fastest along  $\Delta \nu_2=-\Delta \nu_1$, and the region where 
$|\Delta \nu_2 - \Delta \nu_1| > B_{\rm bw}$ is not sampled at all. The right panel, which incorporates the periodic pattern of missing channels present in the MWA data, is very similar to the left panel. We note that the flagging introduces an extra modulation in  $N_\nu(\Delta\nu_1,\Delta\nu_2)$  relative to the left panel. There is also an overall reduction in the sampling due to the missing channels. However, we note that there are no missing frequency separations $(\Delta \nu_1,\Delta \nu_2)$ due to the missing frequency channels i.e. the sampling $N_\nu(\Delta\nu_1,\Delta\nu_2)$  does not become zero anywhere due to the flagged channels. Our analysis involves carrying out a 2D Fourier transform over the $(\Delta \nu_1,\Delta \nu_2)$  plane to evaluate the 3D BS using (Eq.~\ref{eq:3dbs_mabs}). For ease of computating the Fourier transform,  we restrict the subsequent analysis to $-B_{\rm bw}/2 \le (\Delta\nu_1,\Delta\nu_2) \le B_{\rm bw}/2$, which is the largest square region that avoids the range $|\Delta \nu_2 - \Delta \nu_1| > B_{\rm bw}$ that is not sampled. Note that these limits have been imposed in Eq.~(\ref{eq:dft}) that we have introduced earlier.

 \begin{figure}
\centering
\includegraphics[width=.5\textwidth]{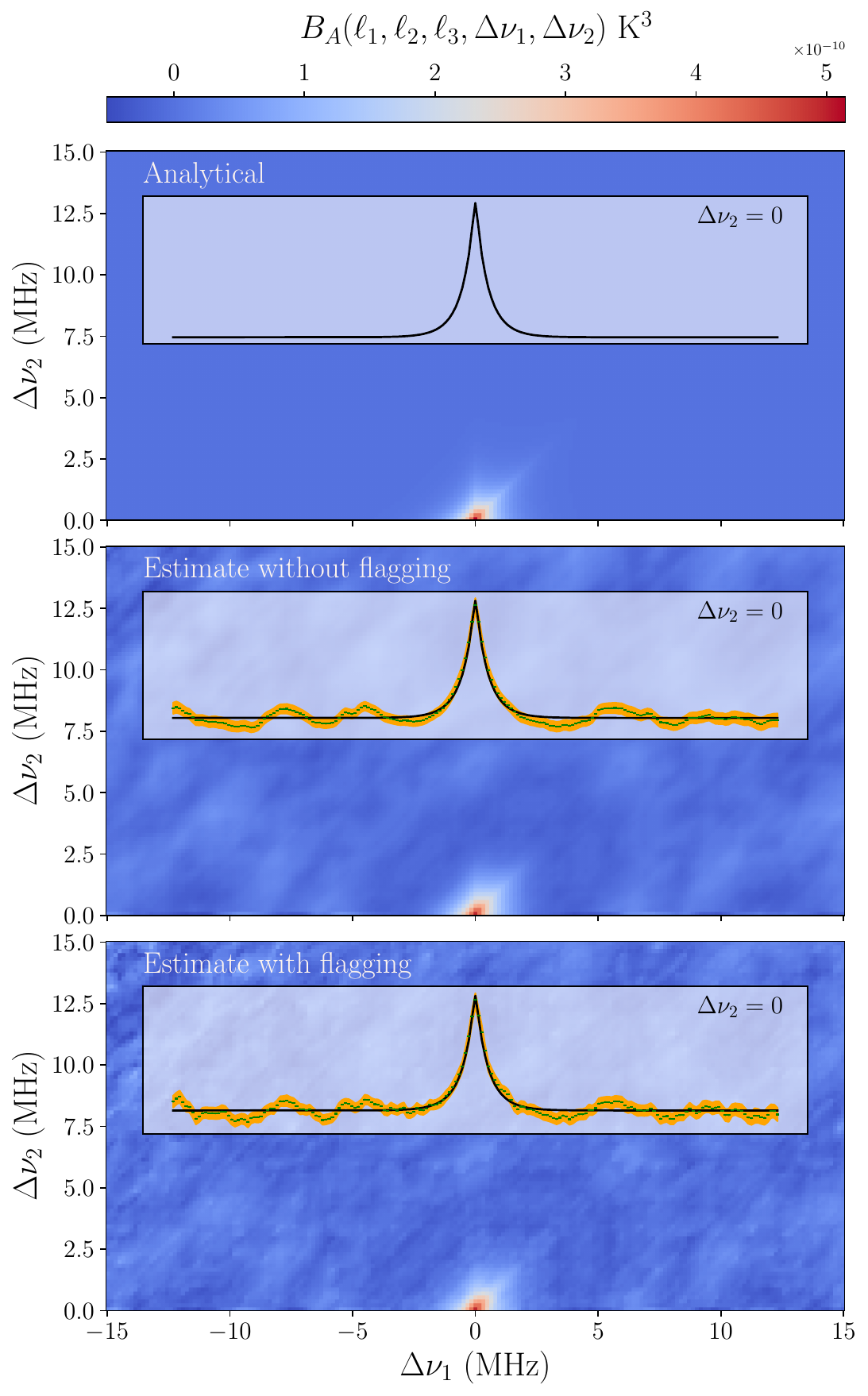}
\caption{MABS $B_A(\ell_1,\ell_2,\ell_3,\Delta \nu_1, \Delta \nu_2)$ as a function of $(\Delta \nu_1, \Delta \nu_2)$ for an equilateral configuration with $\ell_1 = \ell_2 = \ell_3 = 711$. The top panel shows the analytical prediction; the middle and bottom panels show the estimates without and with flagging, respectively. The black solid line in each inset shows a 1D slice of the MABS along $\Delta\nu_1$ at fixed $\Delta\nu_2 = 0$. The green points in the inset show the estimated values, and the orange-shaded region denotes the $1\sigma$ rms statistical fluctuations.}
\label{fig:mabs}
\end{figure}

Fig.~\ref{fig:mabs} presents the  MABS $B_A(\ell_1,\ell_2,\ell_3,\Delta \nu_1, \Delta \nu_2)$ as a function of $(\Delta \nu_1, \Delta \nu_2)$ considering  $\ell_1 = \ell_2 = \ell_3 = 711$ fixed, which corresponds to equilateral triangles in $\bfl$ space. The top panel considers the input model (Eq.~\ref{eq:bsana}), for which it shows the analytical predictions obtained by inverting the Fourier transform in Eq.~(\ref{eq:3dbs_mabs}) and numerically evaluating the integrals. It is evident that the MABS peaks at $\Delta \nu_1 = \Delta \nu_2 = 0$, and de-correlates rapidly as $\Delta \nu_1 $ or $\Delta \nu_2$ increases. The inset panel shows the MABS  as a function of $\Delta \nu_1$ for $\Delta \nu_2 = 0$ fixed. We see that the  MABS de-correlates to $95\%$ of its peak value at $\Delta\nu_2\approx 2$ MHz. 
The de-correlation occurs faster for larger values of $(\ell_1,\ell_2,\ell_3)$ (not shown here). 
This behavior is very similar to the results presented in \citet{Bharadwaj2005} that considers a different model for the input 3D BS. We note in passing that the MAPS $C_{\ell}(\Delta \nu)$ also shows very similar behavior where it peaks at $\Delta \nu=0$, de-correlates rapidly as $\Delta \nu$ increases, and the decorrelation occurs faster if $\ell$ is increased \citep{Bharadwaj_2001,Datta2007}.  The middle and bottom panels, respectively, show the MABS estimated from the simulated visibilities without and with flagging.  All three panels look very similar in the central region, i.e., in the vicinity of $\Delta \nu_1= \Delta \nu_2=0$ where the MABS peaks. Away from this region, the analytical predictions are close to zero, while the estimated MABS shows fluctuations that are visible in the $(\Delta \nu_1, \Delta \nu_2)$ plane. 
These fluctuations are also visible in the respective inset panels where the green points show the mean estimated values, and the orange-shaded region indicates the $1 \sigma$  error bars. The fluctuations in the estimated values arise from statistical fluctuations due to the finite sampling $(N_\nu)$, and we have larger fluctuations when flagging is included.  We see that all the fluctuations are within $\pm 3 \sigma$ from the analytical predictions, which validates our simulations and the MABS estimator.

 \begin{figure*}
\centering
\includegraphics[width=1\textwidth]{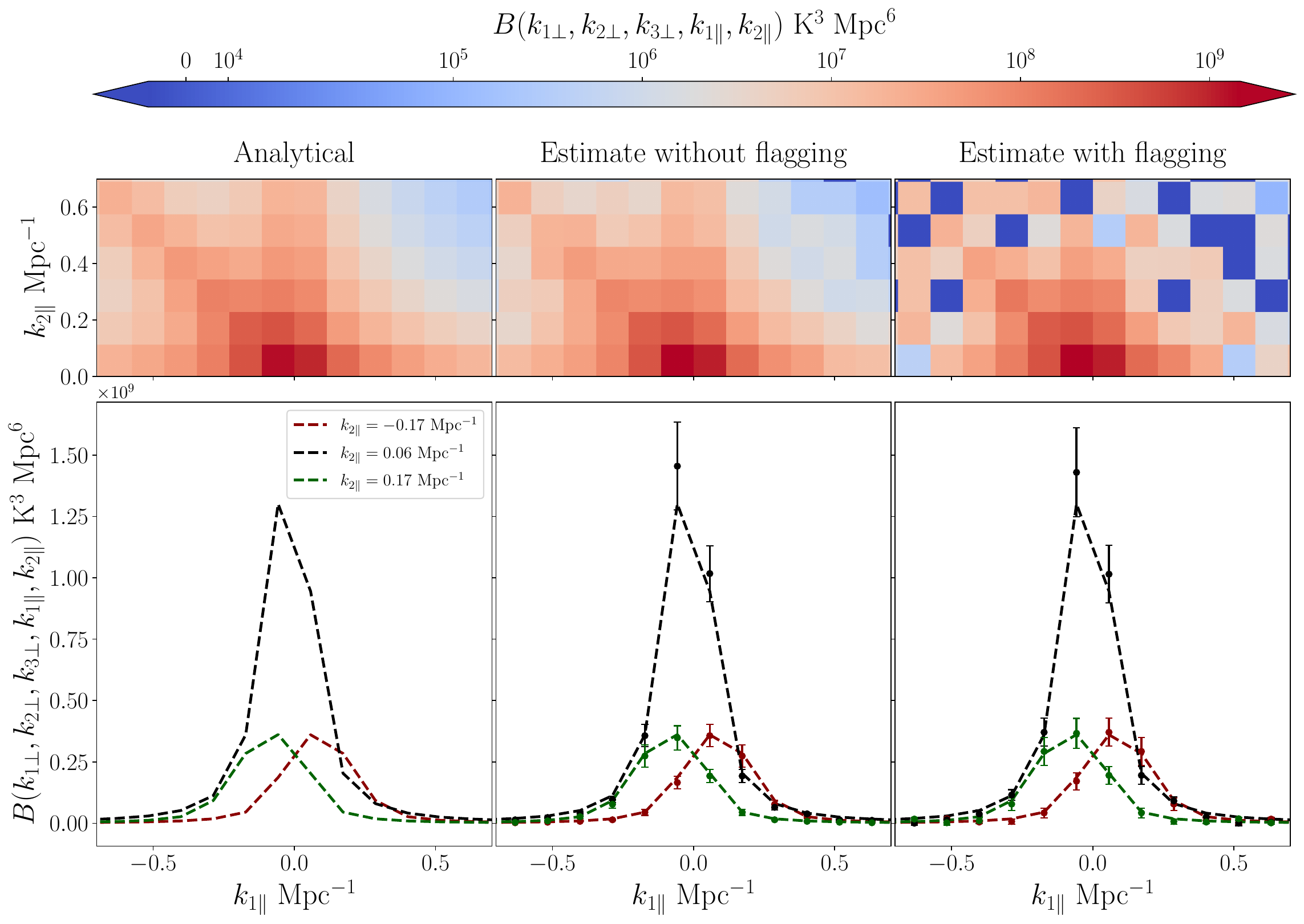}
\caption{ The 3D cylindrical BS $B(k_{1\perp}, k_{2\perp}, k_{3\perp}, k_{1\parallel}, k_{2\parallel})$   with $k_{1\perp} = k_{2\perp} = k_{3\perp} = \frac{\ell_1}{r}= 0.077~\mathrm{Mpc}^{-1}$ fixed. The analytical predictions in the left panels are calculated using Eq.~(\ref{eq:bsana}), whereas the estimates are obtained via a 2D Fourier transform (Eq.~\ref{eq:dft}) of the MABS shown in the respective panels of Fig.~\ref{fig:mabs}. In all the panels, the values have been binned in a $20 \times 20$  grid in the $ (k_{1\parallel}, k_{2\parallel})$  plane. The top row shows the 3D BS as a heat-map in the $(k_{1\parallel}, k_{2\parallel})$ plane,  the bottom row shows 1D slices along $k_{1\parallel}$ for three fixed values of $k_{2\parallel} = \{-0.17, 0.06, 0.17\}~\mathrm{Mpc}^{-1}$, indicated by brown, black, and green lines, respectively. The middle and right panels show the estimates without and with flagging, respectively. The error bars represent the $1\sigma$ rms statistical uncertainty.
 }
\label{fig:cylbs}
\end{figure*}

Fig.~\ref{fig:cylbs} shows ${B}(k_{1 \perp},k_{2 \perp},k_{3 \perp},k_{1\parallel },k_{2\parallel})$  as a function of $(k_{1\parallel },k_{2\parallel})$,  which is 
the 3D  cylindrical BS corresponding to the MABS shown in Fig.~\ref{fig:mabs}. This has been obtained by performing the 2D DFT from $(\Delta \nu_1,\Delta \nu_2)$ to $(k_{1\parallel },k_{2\parallel})$ as given by Eq.~(\ref{eq:dft}). Here,  we have $k_{1\perp}=k_{2\perp}=k_{3\perp}=\ell_1/r = 0.077 {~\rm Mpc}^{-1}$ fixed, which corresponds 
to an equilateral triangles in the $k_{\perp}$ plane. The value of $(k_{1\parallel}, k_{2\parallel})$ span the range $[-1.149~\rm{Mpc}^{-1}, 1.149~\rm{Mpc}^{-1}]$ with a spacing of 
$\Delta k_{\parallel}=0.0124~\rm{Mpc}^{-1}$.  To enhance the signal-to-noise ratio of the estimates, we further bin the results onto a $20 \times 20$ grid in the $(k_{1\parallel}, k_{2\parallel})$ plane. The analytical predictions, shown in the left panels, we calculated on the $(k_{1\parallel },k_{2\parallel})$ grid using Eq.~(\ref{eq:bsana}), and binned in exactly the same way as the estimated values. Note that ${B}(k_{1 \perp},k_{2 \perp},k_{3 \perp},k_{1\parallel },k_{2\parallel})={B}(k_{1 \perp},k_{2 \perp},k_{3 \perp},-k_{1\parallel },-k_{2\parallel})$, and it suffices to show the results for only the upper-half plane. 
 The top row shows a heat-map of $B(k_{1\perp}, k_{2\perp}, k_{3\perp}, k_{1\parallel}, k_{2\parallel})$ as a function of $(k_{1\parallel}, k_{2\parallel})$. The left panel shows the analytical predictions for the BS (Eq.~\ref{eq:bsana}). The BS peaks at  $k_{1\parallel}= k_{2\parallel}\approx 0$ and declines (because $P(k)\propto k^{-2}$)  with increasing $k_{1\parallel}, k_{2\parallel}$. The middle and right panels show the BS estimated without and with flagging, respectively. In both cases, visually, the estimated results agree well with the analytical prediction. However, when flagging is included, we find considerable statistical fluctuations at large $(k_{1\parallel}, k_{2\parallel})$ where the BS has a small value. 
The bottom row show 1D cuts of the BS as a function of $k_{1\parallel}$ for three different values of $k_{2\parallel \,} \in \{-0.17 ~(\rm red),\, 0.06 ~(\rm black),\, 0.17 ~(\rm green)\}\,\text{Mpc}^{-1}$. The solid lines show the analytical predictions, and the bullet points in respective colors show the estimated values with $1\sigma$ error bars. The estimates without and with flagging are both consistent with the analytical predictions. 
The statistical fluctuations are slightly larger due to the reduced sampling when flagging is included.

 \begin{figure*}
\centering
\includegraphics[width=1\textwidth]{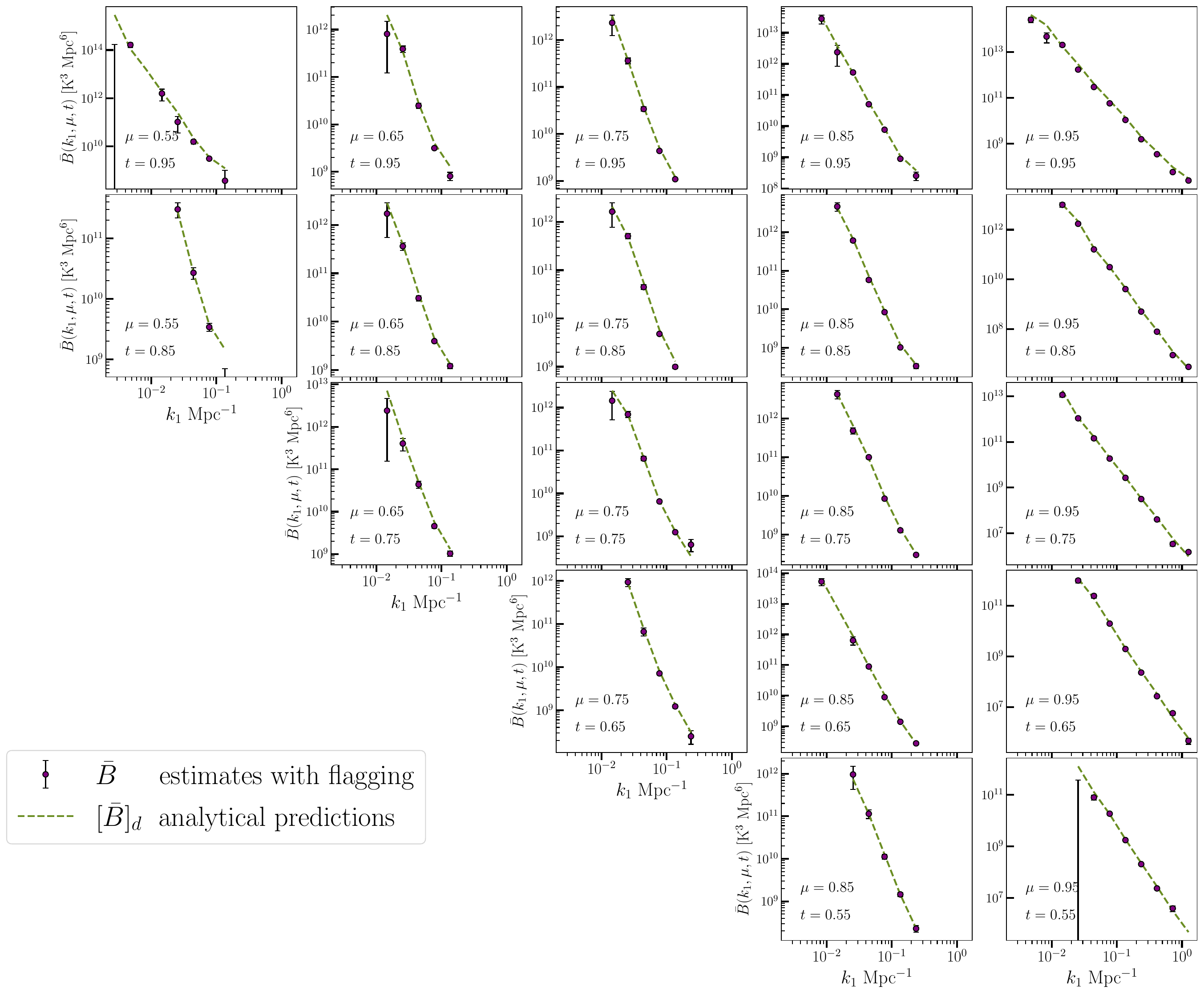}
\caption{The 3D BS monopole $\bar{B}(k_1, \mu, t)$ for all the triangle configurations where we have estimates. Each panel,  which corresponds to a different $(\mu,t)$ (shape),  shows the estimated value $\bar{B}$, with $1 \sigma$ error bars, as a function of $k_1$. The analytical predictions $[\bar{B}]_d$ are also shown for comparison. Here $\mu$ and $t$ respectively vary along the horizontal and vertical directions, exactly as in Figure~\ref{fig:k1_comb}. }
\label{fig:3dbs}
\end{figure*}

Finally, we now turn our attention to the multipoles of the 3D BS. Although, in principle, we can calculate all the multipoles from the cylindrical BS using Eq.~(\ref{eq:bs_lm}), in this work we restrict our analysis to the  monopole  $\bar{B}_0^0(k_1, \mu, t)$, which we denote using 
 $\bar{B}(k_1, \mu, t) \equiv  \bar{B}_0^0(k_1, \mu, t)$ in the rest of this paper. To keep the computational requirement under control, we only consider the first $16$ annular rings, which restricts the analysis to triangle configurations with $k_\perp \le 0.0753~\mathrm{Mpc}^{-1}$. 
Further,  to enhance the signal-to-noise ratio and facilitate visualization of the results, we further bin the $\bar{B}(k_1, \mu, t)$ estimates in the 3D $(k_1, \mu, t)$ parameter space. Here,  
we use 12 logarithmically spaced bins in $k_1$, and 5 linearly spaced bins in  $\mu$ and $t$ each. Note that the estimates of $\bar{B}$ are inherently sampled over a discrete set of $k$-modes that are subjected to binning. However, the theoretical framework Eqs.~(\ref{eq:bsana}), and (\ref{eq:bs_lm}) assumes a continuum of $k$ modes. To account for the effects of discretization and binning, and to ensure a consistent comparison, we evaluate the analytical $\bar{B}$ at the same set of discrete $k$-modes used in the estimates and apply an identical binning procedure. In the subsequent discussion, we refer to the resulting binned analytical predictions as $[\bar{B}]_d$.

Fig.~\ref{fig:3dbs} shows $\bar{B}(k_1, \mu, t)$ as a function of $k_1$ for triangles of all possible shapes. Each panel corresponds to a different shape, as quantified by the 
shape parameters $0.5 \le (\mu,t) \le 1$, which are constrained by the condition $2\mu t \geq 1$. The reader is referred to Fig.~2 of \cite{bharad2020} for a detailed discussion of the parameterization, which we summarize here.  The upper-left corner of the $(\mu, t)$ plane corresponds to equilateral triangles. The upper (lower) boundary  represents  L-isosceles 
(S-isosceles) triangles where the two larger (smaller) sides are equal.  The right boundary corresponds to linear (degenerate) triangles where the three sides are nearly collinear. The upper and lower right corners respectively represent squeezed triangles $(k_1 \approx k_2,\, k_3 \rightarrow 0)$ and stretched triangles $(k_1/2 \approx k_2 \approx k_3)$, respectively. The different panels together allow for a comprehensive examination of the 3D BS across a wide variety of triangle shapes and sizes.

We first consider the top left panel that shows $\bar{B}(k_1,\mu,t)$ as a function of $k_1$ for fixed  $(\mu,t)=(0.55,0.95)$ that corresponds to equilateral triangles.  The filled circles with error bars show the mean and standard deviation of $\bar{B}$ estimated from $250$ realizations of the simulated data with flagging. The results without flagging are very similar and have not been shown here. The green dashed green line shows $[\bar{B}]_d$ the binned analytical predictions. We see that the estimated values are in good agreement with the analytical predictions. Considering the other panels that correspond to different shapes, we see that the accessible $k_1$ range varies with the triangle shape,  and the squeezed triangles demonstrate the largest extent in $k_1$. In general, we see that the estimated values are in good agreement with the analytical predictions across the entire range of $(k_1,\mu,t)$. However, there possibly are a few exceptions at the smallest $k_1$ where we find deviations that could arise because the convolution with the primary beam pattern becomes important in Eq.~(\ref{eq:S3}). This is expected to be important for the smallest annular ring that corresponds to $k_{\perp}=0.002 \, {\rm Mpc{}^{-1}}$ for which the angular scale (set by $\ell=22$) becomes comparable to the FWHM of the primary beam.   The subsequent results provide a more quantitative comparison between  $\bar{B}(k_1, \mu, t)$  with  $[\bar{B}]_d(k_1, \mu, t)$.



 \begin{figure*}
\centering
\includegraphics[width=1\textwidth]{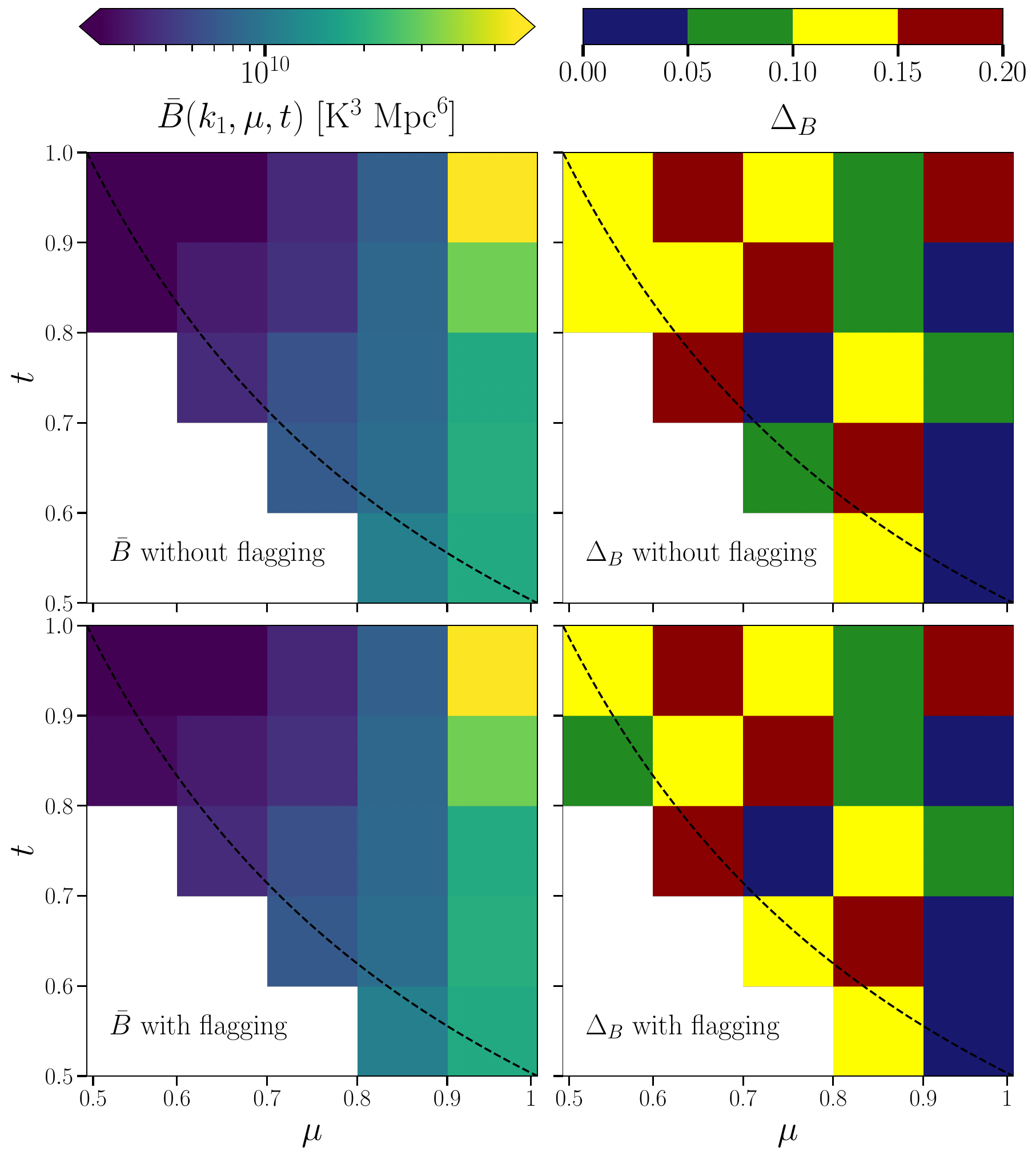}
\caption{ The left column shows the estimated 3D BS monopole $\bar{B}(k_1, \mu, t)$ as a function of $(\mu, t)$ at fixed $k_1=0.078$ Mpc$^{-1}$. The allowed values of $(\mu,t)$, which quantifies the triangle shape,  satisfy the constraint  $2\mu t \geq 1$, indicated by a black dashed line.   The right panels show the corresponding $\Delta_{B} = |\bar{B} - [\bar{B}]_d| / [\bar{B}]_d$, which 
quantifies the fractional deviations relative to  $[\bar{B}]_d$ the analytical predictions.}
\label{fig:k1_comb}
\end{figure*}

The left panels of Fig.~\ref{fig:k1_comb} show the shape ($\mu,t$) dependence of the estimates $\bar{B}$ at fixed size $k_1=0.078\,\mathrm{Mpc}^{-1}$ for without flagging (top-left) and with flagging cases (bottom-left). We see that the variation of $\bar{B}$ in the $\mu-t$ plane is very similar in both cases. We do not show the plot for $[\bar{B}]_d$ since it is visually indistinguishable from $\bar{B}$. The $\bar{B}$ has the largest value for the squeezed triangles ($\mu,t\rightarrow1$). In general,  $\bar{B}$ falls as both $\mu$ and $t$ decrease. The values decrease relatively faster along $\mu$  compared to $t$. Overall, $\bar{B}$ is large for the linear triangles ($\mu\rightarrow 1$), and smallest for the equilateral triangles ($\mu\rightarrow0.5,t\rightarrow1$). Since $\bar{B}(k_1, \mu, t) \propto k_1^{-4}$, we expect the $(\mu, t)$ dependence to remain similar across different values of $k_1$, except for variations arising from binning and discrete sampling effects. The right panels show the fractional deviation, $\Delta_{B}=|\bar{B}-[\bar{B}]_d|/[\bar{B}]_d$, which highlights the relative difference between the estimated and the analytically predicted values. The behavior of $\Delta_{B}$ is, by and large, similar without (top-right)  and with (bottom-right) flagging. We find that in most cases the fractional deviations are within $15\%$, except for $5$ bins where the fractional deviation has values $\approx 15\%$–$20\%$. In summary, the estimated values are in close agreement with the analytical predictions.

 \begin{figure*}
\centering
\includegraphics[width=1\textwidth]{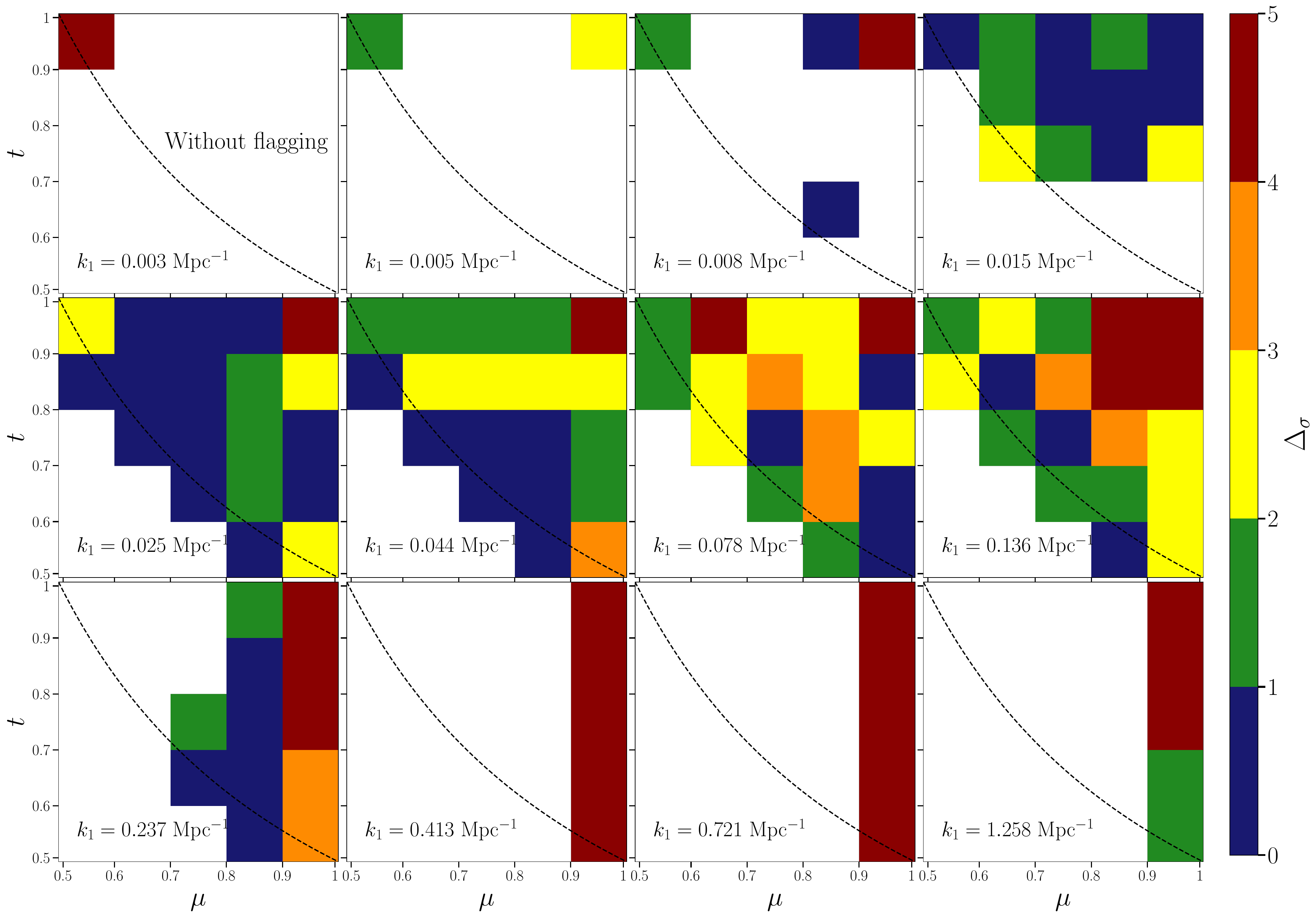}
\caption{Considering all the  estimated 3D BS monopole $\bar{B}(k_1, \mu, t)$  without flagging, this 
shows  $\Delta_\sigma = |\bar{B} - [\bar{B}]_d / \sigma$, which is the deviation  from $[\bar{B}]_d$  the analytical prediction, relative to $\sigma$  the expected statistical fluctuations. Each panel considers a different $k_1$.  We have estimates for only a few  $(\mu,t)$ (shapes) at some $k_1$,   whereas for others,  the entire allowed range ($2\mu t \ge  1$ black dashed line) is available. }  
\label{fig:snr_nf}
\end{figure*}

 \begin{figure*}
\centering
\includegraphics[width=1\textwidth]{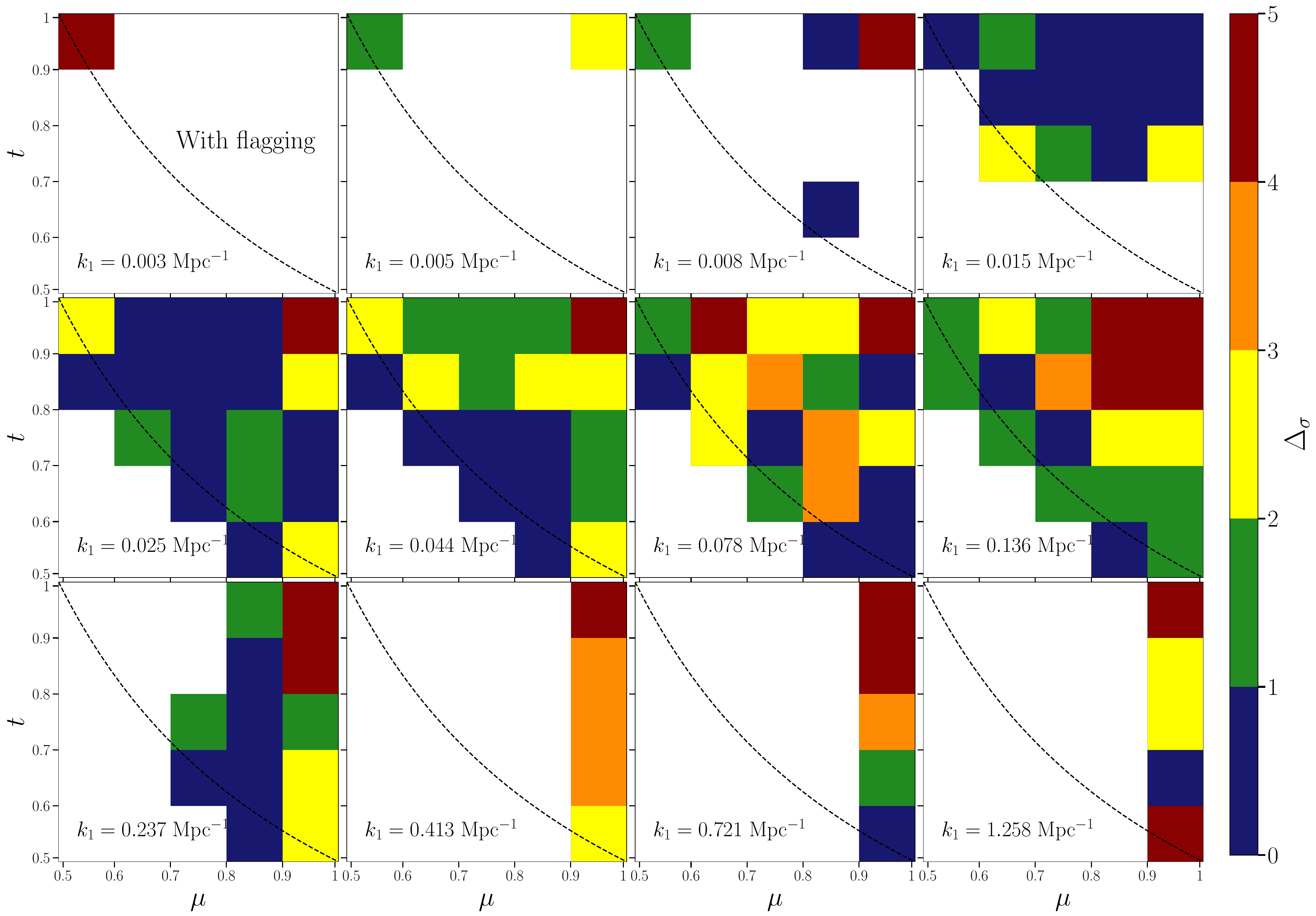}
\caption{Same as Fig.~\ref{fig:snr_nf}, but with flagging.}
\label{fig:snr_f}
\end{figure*}


To assess whether the deviations between the estimated values $\bar{B}(k_1, \mu, t)$ and the analytically predicted values $[\bar{B}]_d$ are consistent with statistical fluctuations, we consider the ratio $\Delta_\sigma = |\bar{B} - [\bar{B}]_d| / \sigma$, which is shown in Figures~\ref{fig:snr_nf} and~\ref{fig:snr_f},  without and with flagging, respectively. The results are presented for all 12 $k_1$ bins, covering the range $[0.003~\mathrm{Mpc}^{-1}, 1.258~\mathrm{Mpc}^{-1}]$, with each panel showing the $\Delta_\sigma$ as a function of $(\mu, t)$ for a specific $k_1$ bin. Note that the plots cover all the combinations of $(k_1,\mu,t)$ for which the estimates of $\bar{B}$ are presented in Fig.~\ref{fig:3dbs}. Considering the smallest $k_1 (=0.003 ~\mathrm{Mpc}^{-1})$, we have an estimate for the equilateral configuration only, for which $4<\Delta_\sigma\leq5$ for both without and with flagging. This relatively large deviation can be attributed to the convolution with the primary beam at large angular scales, as discussed earlier. As $k_1$ increases, the coverage in the $(\mu, t)$ space gradually expands, starting with the squeezed and L-isosceles triangles.  For the range $0.005 \leq k_1 \leq 0.015~\mathrm{Mpc}^{-1}$, $\Delta_\sigma$ remains below 3 for most configurations, except for squeezed triangles at $k_1 = 0.008~\mathrm{Mpc}^{-1}$, where $4< \Delta_\sigma\leq 5$. At higher $k_1$ bins, $0.025 \leq k_1 \leq 0.136~\mathrm{Mpc}^{-1}$, the estimates are available across the full $(\mu, t)$ domain. For most configurations in this range, $\Delta_\sigma \leq 3$, and a few bins show slightly higher deviations in the range $3 < \Delta_\sigma \leq 4$.  Notably, the squeezed configurations show large deviations with $4<\Delta_\sigma\leq 5$. This can be attributed to the fact that this bin incorporates triangles with very small $k_3$, and the value of the BS (Eq.~\ref{eq:bsana})  changes
very rapidly even within the bin (see \citealt{gill_2024} for a detailed discussion). Considering the next bin $k_1=0.237~\mathrm{Mpc}^{-1}$, the coverage in the $(\mu, t)$ space declines, with estimates available for roughly half the $\mu-t$ space (primarily for $\mu \gtrsim 0.75$). In this bin, $\Delta_\sigma \leq 3$ for many configurations, but again increases to $4 < \Delta_\sigma \leq 5$ near the squeezed limit. Considering  larger $k_1$  $[ 0.413~\mathrm{Mpc}^{-1},1.258~\mathrm{Mpc}^{-1}]$, we obtain estimates for linear triangles only. The deviations lie in the range $4 < \Delta_\sigma \leq 5$ for most estimates in the case without flagging. In contrast, with flagging, the values of $\Delta_\sigma$ are reduced, and only a few bins near the squeezed limit show such high deviations.  Considering the entire $(k_1,\mu,t)$ space,  the values of $\Delta_\sigma $   indicate that the estimated values of the BS are largely consistent with the analytical predictions, and in most cases the deviations between the two are consistent with the expected statistical fluctuations. In some cases, particularly for small $k_1$ or near the squeezed limit, we have relatively large deviations ($4 < \Delta_\sigma \le 5$). However, in all the cases we have $\Delta_\sigma < 5$, which does not exclude the possibility that these deviations arise from statistical fluctuations.

\section{Summary and Conclusion}
\label{sec:sum}
It is believed that hydrogen reionization occurred through the onset and growth of ionized bubbles, and the resulting EoR 21 cm signal is predicted to be highly non-Gaussian \citep{Bharadwaj2005}. 
 Radio interferometric observations have primarily focused on the 21 cm PS to quantify the statistics of the EoR 21 cm signal, however, this is insensitive to non-Gaussianity. 
The 21 cm BS offers a complementary probe that is sensitive to non-Gaussianity.   This is also sensitive to the geometry and topology of the \Hi~ distribution. 
 
In this paper, we build upon an earlier work \citetalias{Gill_2024_2d3vc}, which presented an estimator for the angular BS considering single-frequency radio-interferometric observations. Here, we consider multi-frequency radio observations and present a fast and efficient visibility-based estimator for both the MABS and the 3D 21 cm BS. The three visibility correlations at different frequencies are directly related to the MABS. Considering the additional assumption of ergodicity along frequency, the cosmological 21 cm BS is obtained through a 2D Fourier transform of the MABS. The computation required to directly correlate the visibilities scales as the cube of the number of visibilities and frequency channels, making it computationally prohibitive. To address this, we work with gridded visibilities, which reduce the complexity, but a naive implementation still scales as $N_c^3 N_t^4$, where $N_c$ is the number of frequency channels and $N_t$ is the number of grid points. To further accelerate the computation, we incorporate the FFT-based approach \citep{Sefusatti_thesis,Jeong_thesis,sco2015}, which is predicted to scale as  $N_c^3 N_t^2\log(N_t^2)$. For the analysis presented in this work, the full computation in a single run takes approximately one hour on a 16-core CPU.

We validate our estimator using simulated data for a single pointing of the MWA drift scan  observations \citep{Patwa_2021}, where the non-Gaussian sky signal is generated from a model with a known input BS. The data spans a total bandwidth of $30.72$ MHz, centered at $154.25$ MHz. We carried out the analysis for two cases: one without flagging, where no data were flagged, and another with flagging, where the simulated data were flagged for the frequency channels and baselines corresponding to actual observations. We find that, in both cases, the MABS and the 3D BS can be reliably estimated across a wide range of triangle configurations, covering the scales $0.003 ~\mathrm{Mpc}^{-1}\leq k_1 \leq 1.258  ~\mathrm{Mpc}^{-1}$. The estimated BS values show excellent agreement with the model predictions, with deviations that are consistent with the expected statistical fluctuations. These results confirm the accuracy and robustness of our estimator. Despite the loss of data from flagging, the estimator is able to recover the BS with deviations less than $20\%$. 

The analysis presented here does not include real observational effects such as astrophysical foregrounds, system noise, and instrumental systematics. In future work, we aim to apply our estimator to the simulated foreground signal and include the system noise contribution. Work is currently underway to apply the estimator to the actual MWA data.

\section*{Acknowledgements}
We acknowledge the computing facilities in the Department of Physics, IIT Kharagpur.

\section*{Data Availability}

The simulated data and the package involved in this work
will be shared on reasonable request to the authors.

\bibliography{main}{}
\bibliographystyle{aasjournal}


\label{lastpage}
\end{document}